\documentclass[12pt]{article}
\usepackage{graphicx}

\begin{document}

\begin{center}

\Large{Diminished Upper Bounds on the Unification Mass Scales for Heavy Higgs Boson Masses}

\end{center}

\begin{center}
\large{V. Elias, S. Homayouni and D. J. Jeffrey}

\smallskip

{\small \it Department of Applied Mathematics,\\ The University of Western Ontario\\London, Ontario N6A 5B7 CANADA}

\end{center}

\vskip 1cm

\abstract{We consider dominant 3-, 4-, and 5-loop contributions to $\lambda$, the quartic scalar coupling-constant's $\beta$-function in the Standard Model.  We find that these terms accelerate the evolution of $\lambda$ to nonperturbative values, thereby lowering the unification bound for which scalar-couplings are still perturbative.  We also find that these higher order contributions imply a substantial lowering of $\lambda$ itself before the anticipated onset of nonperturbative physics in the Higgs sector.}

\vskip 1cm

\baselineskip 24pt

The dominant running coupling constants of the standard model evolve with $\mu$, the renormalization scale, according to 2-loop renormalization group equations
\begin{eqnarray}
\lefteqn{\mu\frac{d\lambda}{d\mu}=\frac{1}{16\pi^2}\{4\lambda^2+12\lambda h^2-36h^4-9\lambda g^2_2-\frac{9}{5}\lambda g_1^2+\frac{81}{100}g_1^4 
}\nonumber\\
&&{}+\frac{27}{10}g_1^2g_2^2+\frac{27}{4}g_2^4\}+\frac{1}{(16\pi^2)^2}\{-\frac{26}{3}\lambda^3-24\lambda^2 h^2-3\lambda h^4\nonumber\\
&&{}+180h^6+80\lambda g_3^2h^2-192h^4g_3^2+{}\cdots \}\\
\lefteqn{\mu\frac{d h}{d\mu}=\frac{1}{16\pi^2}\{\frac{9}{2}h^3-8g_3^2h-\frac{9}{4}g_2^2h-\frac{17}{20}g_1^2h \}}\nonumber\\
&&{}+\frac{1}{(16\pi^2)^2}\{-12h^5-2\lambda h^3+\frac{1}{6}\lambda^2h\nonumber\\&&{}+36g_3^2h^3-108g_3^4h+{}\cdots \}\\
\lefteqn{\mu\frac{d g_3}{d\mu}=\frac{1}{16\pi^2}\{-7g_3^3 \}}\nonumber\\
&&{}+\frac{1}{(16\pi^2)^2}\{-26g_3^5-2h^2g_3^3+\frac{11}{10}g_1^2g_3^3+\frac{9}{2}g_2^2g_3^3 +{}\cdots\}\\
\lefteqn{\mu\frac{d g_2}{d\mu}=\frac{1}{16\pi^2}\{\frac{-19}{6}  g_2^3 \}}\nonumber\\
&&{}+\frac{1}{(16\pi^2)^2}\{\frac{35}{6}g_2^5+12g_3^2g_2^3+\frac{9}{10}g_1^2g_2^3-\frac{3}{2}h^2g_2^3 +{}\cdots\}\\
\lefteqn{\mu\frac{d g_1}{d\mu}=\frac{1}{16\pi^2}\{\frac{41}{10}  g_1^3 \}}\nonumber\\
&&{}+\frac{1}{(16\pi^2)^2}\{\frac{199}{50}g_1^5+\frac{27}{10}g_2^2g_1^3+\frac{44}{5}g_3^2g_1^3-\frac{17}{10}g_1^3h^2 +{}\cdots\}
\end{eqnarray}
In the above equations the initial conditions for the gauge coupling constants $g_3$, $g_2$ and $g_1$ are obtained from low-energy phenomenology $[ \alpha_s (M_z) = 119,\; \;  \alpha(M_z) = 1/128,  \; \; sin^2 \theta_w = 0.225]$.  The top quark mass leads to a numerical initial value for the Yukawa coupling constant $h(\mu)$.  These numerical initial conditions are
\begin{eqnarray}
g_1 (M_z) = 0.4595,\nonumber\\
g_2 (M_z) = 0.6605,\nonumber\\
g_3 (M_z) = 1.2228,\nonumber\\
h (M_t) = 1.0020.
\end{eqnarray}
Only $\lambda$ has an unspecified initial condition.  The initial value for $\lambda$ may be expressed in terms of the Higgs boson mass
\begin{equation}
\lambda (M_H) = 3 M_H^2 / v^2
\end{equation}
where $v = 246$ GeV is the electroweak vacuum expectation value.  Thus a large Higgs boson mass necessarily implies a large value of $\lambda (M_H)$ that will evolve to increasingly large values of $\lambda(\mu)$ as $\mu$ increases.

The idea that the scalar interaction (as well as all other standard model interactions) remain perturbative up to unification \cite{2} necessarily implies an upper bound on the unification mass scale $M$ for a given choice of $M_H$ by the requirement $\lambda (M) = \lambda_{max}$, where $\lambda_{max}$ is the largest value of $\lambda$ for which scalar field theory should remain perturbative.  This criterion has been used by Riesselmann and collaborators \cite{3} to correlate upper bounds on $M$ with $M_H$.

In the present note we reassess these bounds by considering the purely scalar-field (i.e., purely $\lambda$) 3- 4- and 5-loop contributions to the $\beta$-function (1).  These have been known for some time;  the scalar field theory projection of the standard model is just a globally $O(4)$-symmetric real scalar field theory whose $\beta$-function is given to 5-loop order by \cite{4}
\begin{eqnarray}
\mu\frac{d}{d\mu} Y & = & 4Y^2 - \frac{26}{3} Y^3 + 55.661 Y^4 - 532.99 Y^5 + 6317.7 Y^6 \ldots\nonumber\\
& & \left( Y \equiv \lambda/16\pi^2\right)
\end{eqnarray}
This expression has a bearing both on how $\lambda_{max}$ is obtained, as well as how rapidly $\lambda$ itself evolves to $\lambda_{max}$.

Prior calculations of the upper bound of the unification mass scale assumed $\lambda_{max}$ was equal (or related to) its ``fixed-point'' value $\lambda_{FP}$, defined as where the two loop and one loop contribution to (8) are equal:
\begin{eqnarray*}
Y = 6/13, \; \; {\rm or} \; \;  \lambda_{FP} \cong 73
\end{eqnarray*}
In a two loop world, this would be near a fixed point in the RG equation (1), particularly as $\lambda$ is so dominant a coupling constant compared to the others in Eq. (1).  In fact, people have advocated for various reasons that $\lambda_{max}$ be $\lambda_{FP} / 2$ \cite{5} or even smaller \cite{6}.  The top curve of Fig. 1 shows, given a choice of $M_H$, the corresponding value of the upper bound $M$ for the unification mass scale, given that $\lambda_{max} = \lambda_{FP} / 2 = 36.$  The intermediate curve in Fig. 1 shows for $\lambda_{max} = \lambda_{FP} / 2$ how the upper-bound $M$ on the unification mass scale decreases if the 3-, 4-, and 5-loop terms in Eq. (8) are incorporated into the $\lambda$ $\beta$-function (1).

However, the additional $\beta$-function terms in (8) make any referencing to $\lambda_{FP}$ irrelevant.  The $\beta$-function series (8) does not monotonically decrease unless $Y < 0.084$ ($\lambda < 13.3$).  Hence $\lambda_{max} = 13.3$ is an {\it upper bound} on the value of $\lambda$ for which perturbative Higgs sector physics may still be possible, in that 4 and 5 loop terms in (8) are equal. The evolution of the coupling constant $\lambda$ should also be inclusive of the 3-, 4-, and 5-loop terms of Eq. (8), as in the middle curve, since such terms are  comparable when $\lambda_{max} = 13.3$.  When we augment Eq. (1) with these 3-5 loop terms in Eq. (8), and impose the additional requirement that the upper bound on $\lambda$ for perturbative physics is 13.3, we obtain the lowest of the three curves in Fig. 1.

Fig. 1 shows that a given value for $M$, the upper bound for the unification mass scale, now corresponds to substantially smaller values of the Higgs mass when Eq. (8) augments the Eq. (1) $\beta$-function, and when $\lambda_{max} = 13.3$.  This separation becomes pronounced when $M < 10^5$ GeV.  By incorporating Eq. (8), we find that a Higgs mass of $304$ GeV can occur in a theory only if unification is prior to $100$ TeV;  a Higgs mass of $360$ GeV can occur only if unification is prior to $10$ TeV; and that Higgs mass in excess of $460$ GeV would involve non-perturbative physics immediately.  In the prior analysis (top curve) this same non-perturbative bound would be in excess of $800$ GeV.

We reiterate that the reduction we find in the unification mass-scale upper bound $M$ for a given choice of $M_H$ is itself conservatively taken.  The choice $\lambda_{max} = 13.3$ assumes perturbative physics even when 3-, 4-, and 5-loop contributions to the $\beta$-function(8) are comparable in magnitude. One could argue for $\lambda_{max} = 6.65$ via whatever reasoning already employed in the past for choosing $\lambda_{max} = \lambda_{FP} / 2$ instead of $\lambda_{FP}$. We also note that the effect of higher-than-2 loop contributions becomes unimportant for Higgs masses in the vicinity of $200$ GeV.  We find, for example of a $200$ GeV Higgs boson mass, that the upper bound on the unification mass scale is $10^{12}$ GeV;  for a $190$ GeV Higgs boson mass, the upper bound on the unification mass scale goes up to $10^{15}$ GeV.  In the prior 2-loop analysis $(\lambda_{max} = \lambda_{FP} / 2)$ the same values of the unification mass scale are achieved by Higgs masses only $10$ GeV or so larger than those quoted above.

We are grateful for support from the Natural Sciences and Engineering Research Council of Canada.

\begin{center}
\begin{figure}[htb]
\vspace{9pt}
\includegraphics[angle=360,width=40pc,scale=0.6]{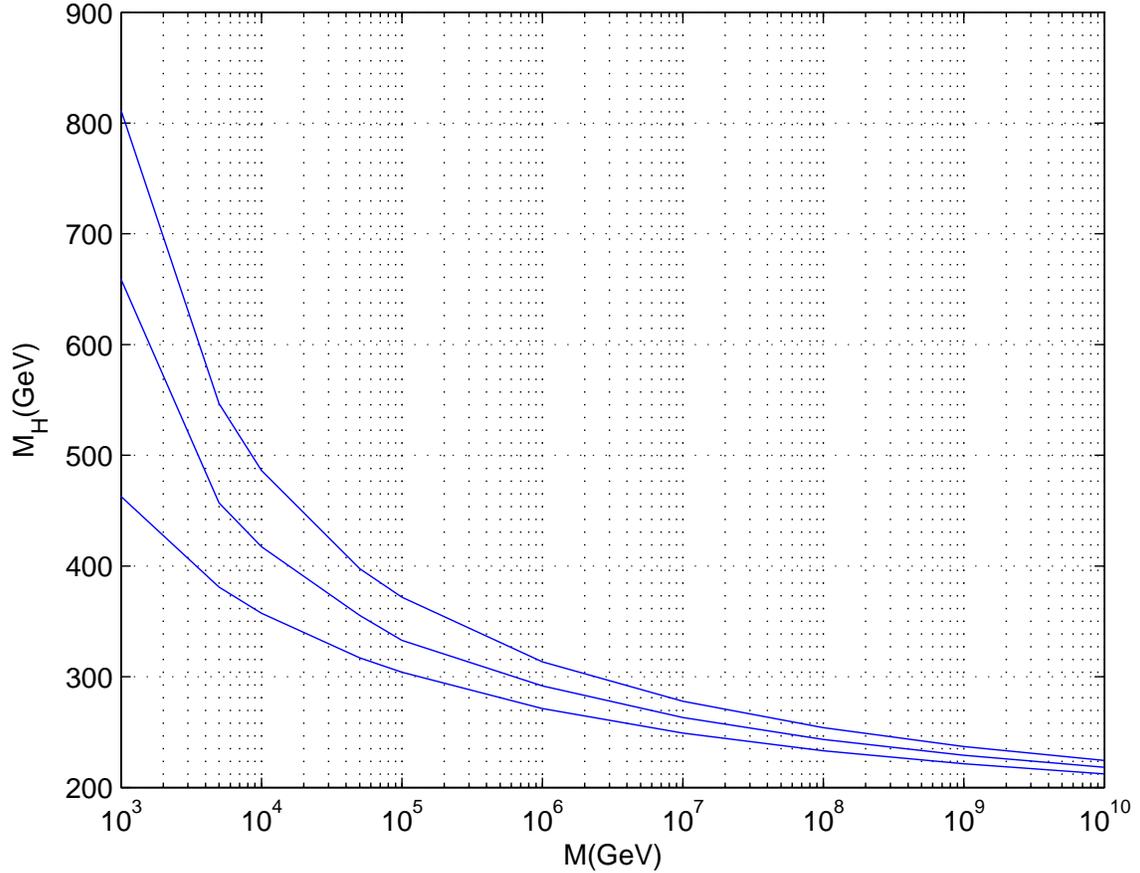}
\caption{Top curve, upper bound on unification mass scale $M$ with no higher-than-2-loop input.  Middle curve, upper bound with 3-5 loop contributions to the $\beta$-function for $\lambda$, but with $\lambda$ assumed perturbative up to $\lambda_{FP} / 2$, as in top curve.  Bottom curve, upper bound with 3-5 loop contributions and concomitant reduction in how large $\lambda$ can be before it is nonperturbative.}
\label{Figure 1}
\end{figure}
\end{center}


\begin{thebibliography}{99}
\bibitem{1}See Appendices of C. Ford, D. R. T. Jones, P. W. Stephenson, M. B. Einhorn, Nucl. Phys. B 395 (1993) 17.
\bibitem{2}V. Elias, Phys. Rev. D 20 (1979) 262.
\bibitem{3}K. Riesselmann, Acta. Phys. Polon. B 27 (1996) 3661.
\bibitem{4}H. Kleinert et al., Phys. Lett B 272 (1991) 39; (E) 319 (1993) 545.
\bibitem{5}T. Hambye and K. Riesselmann, Phys. Rev. D 55 (1997) 7255.
\bibitem{6}K. Riesselmann and S. Willenbrock, Phys. Rev. D 55 (1997) 311.
\end{thebibliography}
\end{document}